\documentclass[a4paper,12pt,draft]{article}

\usepackage[a4paper]{geometry}
\usepackage[utf8]{inputenc} 
\usepackage{hyperref}
\usepackage{float}
\usepackage{amssymb}
\usepackage{amsmath}
\usepackage{amsthm}
\usepackage{url}
\usepackage{amsfonts}
\usepackage{graphicx, float, tabularx}
\usepackage{color}
\usepackage{enumerate}
\usepackage[american]{babel}
\usepackage{stackrel}
\usepackage[numbers,sort,compress]{natbib}
\newcommand{\non}{\nonumber}
\newcommand{\be}{\begin{equation}}
\newcommand{\ee}{\end{equation}}
\newcommand{\bea}{\begin{eqnarray}}
\newcommand{\eea}{\end{eqnarray}}
\newcommand{\mpl}{M_{\rm P}}
\newcommand{\mpr}{m_{\rm p}}
\newcommand{\mel}{m_e}
\newcommand{\qpr}{q_{\rm p}}
\newcommand{\lpl}{L_{\rm P}}
\renewcommand{\d}{\mathrm{d}}

\newcommand{\C}[1]{ {\cal #1}}

\newcommand{\p}{\partial}

\renewcommand{\frac}[2]{{{\displaystyle #1}\over{\displaystyle #2}}}

\author{Piero Nicolini\thanks{E-mail: \texttt{nicolini@fias.uni-frankfurt.de}}
\\[1ex]
\small Frankfurt Institute for Advanced Studies (FIAS)\\[-0.5ex]
\small Ruth-Moufang-Str.~1, D-60438 Frankfurt am Main, Germany\\[1ex]
\small Institut f\"{u}r Theoretische Physik, Goethe-Universit\"{a}t Frankfurt am Main\\[-0.5ex]
\small Max-von-Laue-Str.~1, D-60438 Frankfurt am Main, Germany\\[1ex]
}
\date{}
\title{Planckian charged black holes in  ultraviolet self-complete quantum gravity}

\begin{document}
\maketitle

\vspace{0.1cm}

\begin{abstract}
\noindent  
We present an analysis of the role of the charge within the self-complete quantum gravity paradigm. By studying the classicalization of generic ultraviolet improved charged black hole solutions around the Planck scale, we showed that the charge introduces important differences with respect to the neutral case. First, there exists a family of black hole parameters fulfilling the particle-black hole condition. Second, there is no extremal particle-black hole solution but quasi extremal charged particle-black holes at the best. We showed that the Hawking emission disrupts the condition of particle-black hole. By analyzing the Schwinger pair production mechanism, the charge is quickly shed and the particle-black hole condition can ultimately be restored in a cooling down phase towards a zero temperature configuration, provided non-classical effects are  taken into account. 

\end{abstract}

\thispagestyle{empty}
\newpage




\section{Introduction}
\label{sec:intro}

General relativity (GR) can be considered among the most successful physical theories. It predicts with formidable accuracy the revolution period of binary pulsar systems \cite{WTF81,TaW82,TaW89,WNT10}, and more importantly, it is still getting corroborations by observational data, \textit{e.g.}, the direct detection of gravitational waves \cite{LIGO16,LIGO16b,LIGO17}.  

Despite such a success, GR is not free of problems. It presents limitations in the large distance regime as well as at the shortest length scales \cite{CaD11}. Alternative theories of gravity have been proposed in order to explain the accelerated expansion and the structure formation without invoking dark energy or dark matter components. Quantum gravity, on the other hand, is expected to provide the ultraviolet completion of GR and describe gravity  at Planckian energy scales.

Among all the predictions of GR, black holes are certainly the most sensitive to the gravity short distance behavior. Classical black hole solutions display a curvature singularity and an entropy proportional to the event horizon area in Planck units. From this perspective, every black hole event horizon is a holographic surface pixelized in fundamental cells \cite{Bek73}. Being the black hole temperature inversely proportional to the mass, $T\propto 1/M$, it is natural to consider the case of microscopic black holes to study quantum mechanical effects like the Hawking radiation. Such a breed of black holes might have primordially been produced, via a variety of mechanisms, including the gravitational collapse of the early Universe fluctuations \cite{Haw71,CaH74} or the quantum mechanical decay of deSitter space \cite{MaR95,BoH96,MaN11}. Being extremely hot, microscopic black holes undergo a rapid decay and disclose a further inconsistency of GR, namely the end stage of the evaporation. The latter is plagued by a divergent temperature and cannot longer be described in semi-classical terms.

To solve the problem of black hole spacetimes at short scales, new metrics have been proposed on the basis of quantum gravity arguments. As a general result, one has that quantum mechanical effects can improve the divergent behavior \cite{BoR00,IMN13} or even replace the curvature singularity with a regular spacetime region \cite{Bar68,AyG98,AyG99,AyG99b,AyG00,AyG05,Dym92,Dym02,Dym03,Hay06,BDD03,BrF06,BrD07,BMD07,Mod04,
Mod06,Nic05,NSS06a,NSS06b,ANS07,Nic09,SSN09,NiS10,SmS10,MoN10b,MMN11,Nic12,SpS14,SpS15,SpS17a,SpS17b,NSS17}.

In the context of Planckian black holes there exist even more radical proposals. Rather than assuming gravity as a starting point, one can consider a purely quantum mechanical formulation for a microscopic black hole, provided one recovers the GR description in the large distance/large particle number limit. Such models include the horizon wave function formalism \cite{CaS14,CMS14,CMN15b,CGM16}, the particle-like black hole description \cite{SpS16,SpS17book}, the quantum $N$-portrait \cite{DvGo12,DvG13b,DvG14,DvGo14}, the quantum bound state description \cite{HoR16,GHM15}, the black hole precursor \cite{Cal14,CaC15} and the fuzzball model \cite{Mat05,SkT08}. 
Such pregeometrical models can be roughly grouped under an umbrella paradigm called ``gravity ultraviolet self-completeness''. This term refers to a special character of gravity. Contrary to ordinary non-renornalizable theories, gravity does not admit an obvious separation between a perturbative regime and a non-perturbative one. Gravity is problematic just at the Planck scale. Above the Planck scale, gravity works classically and no quantum gravity description has to be invoked \cite{DGG11,MuN12,AuS13,Car14}. In support of this claim, it can be shown that the scattering of two objects at trans-Planckian energies leads to a classical state, namely a black hole \cite{DFG11,DvG12,DGI15,DGL15}. As a result sub-Planckian distances are no longer accessible and gravity is ultraviolet self-complete \cite{DvG10,DFG12}.  This is equivalent to saying that curvature singularities are just an artifact of the geometric description of gravity. The latter virtually ceases to exist below the Planck length where only a quantum  pregeometry is available. 

The transition between the quantum and the geometrical description is often termed as ``classicalization'' \cite{Dva17}. From a black hole physics perspective, the classicalization produces a black hole that is approximately classical. In \cite{NiS12,FKN16}, an effective metric has been proposed to smoothly interpolate the two opposite regimes. The transition, however, might be non-analytic. In this case,  deviations from classical black hole geometries might be even smaller and negligible around the Planck scale too. The Schwarzschild metric is, however, inadequate to fit in the self-completeness paradigm. One of the major drawbacks is the huge quantum back-reaction the metric suffers when the black hole mass approaches the Planck mass, $M\gtrsim\mpl$. More seriously, the Schwarzschild metric admits event horizons at any length scales, even below the Planck length, $r_{\mathrm{H}}<\lpl$, for $M<\mpl$. Such limitations lead us to consider families of metrics that allow for the horizon extremisation close to the Planck scale. From this vantage point, one can employ the semiclassical approximation up to the Planck scale, being the ratio temperature mass small, $T/M\ll 1$, during the whole black hole life \cite{DeS78,BaB88,BaB89}. Interestingly, the extremal configuration corresponds to a threshold mass, $M_\mathrm{e}$, below which no horizon forms. The degenerate horizon is also a zero temperature state. As a result, the Hawking decay is expected to undergo a SCRAM\footnote{This term, previously known as ``black hole switching off'' \cite{DeS78}, has been adopted in \cite{Nic09} by borrowing the terminology for emergency shutdowns of nuclear reactors.  SCRAM is a backronym for ``Safety Control
Rod Axe Man'', introduced by Enrico Fermi in 1942, during the Manhattan Project at Chicago Pile-1.}, \textit{i.e.}, an asymptotic relaxation at $M\approx M_\mathrm{e}\simeq \mpl$, that precludes the access to length scales smaller than $r_\mathrm{e}=r_\mathrm{H}(M_\mathrm{e})\simeq \lpl$. 

Within such a framework, one might be led to consider the case of Planckian charged black holes as privileged models fitting in the gravity self-completeness paradigm. There exist both technical and phenomenological arguments in support of this choice. First, charged black holes naturally encode the sought horizon extremization. This is related to the instability  of the Schwarzschild metric in the parameter space of solutions. The $r=0$ surface is turned into a timelike surface for any perturbation of the static neutral black hole. Second, it has been shown that Planckian charged black holes might have plentifully been created in the early Universe due to relevant cold, ultracold, lukewarm and Nariai instanton contributions \cite{MaR95}. 

As a result, we present an analysis of the properties of Planckian charged black hole solutions in order to scrutinize the self-complete character of gravity. The paper is organized as follows: in Section \ref{sec:chargedmetric}, we present a general set up for charged black hole solutions in short scale modified gravity and their thermodynamics. In Section \ref{sec:classicalization}, we consider the classicalization of the solutions presented in Section \ref{sec:chargedmetric}. In Section \ref{sec:decay}, we study the evolution of such Planckian black holes while in Section \ref{sec:concl} we draw our conclusions.


\section{Charged black hole metrics at the Planck scale}
\label{sec:chargedmetric}

We start from a generic charged, spherically symmetric black hole metric of the kind 
\begin{equation}
 \d{s}^2=-F(r)\d{t}^2+F^{-1}(r)\d{r}^2+\d{\Omega}^2
 \label{eq:metric}
\end{equation}
where
\begin{equation}
F(r)=1-2G\ \frac{m(r)}{r}+\frac{q^2(r)}{r^2}
\end{equation}
Here the functions $m(r)$ and $q(r)$ 
account for gravity and electrodynamics short scale modifications that are expected to occur in a region of size $\lpl$ around the origin. For larger radii the above metric has to match the usual Reissner-Nordstr\"{o}m geometry. This implies a condition on the profile of the above functions, \textit{i.e.}, for $r\gg\lpl$ one has $m(r)\to M$ and $q(r)\to Q$, where $M$ is the ADM mass and $Q$ is the total charge of the system.

Solutions like \eqref{eq:metric} can be obtained by short scale improved gravity and electrodynamics actions of the kind
\begin{equation}
{\cal S}_\mathrm{tot}=
 \frac{1}{16\pi G}\int \mathfrak{F}({\cal R},\, \Box/\mu^2,\, \dots)\, {\sqrt {-g}}\,\mathrm {d} ^{4}x\; + \int  \mathfrak{L}({\cal F}^2,\, \Box/\mu^2)\, {\sqrt {-g}}\,\mathrm {d} ^{4}x\; ,
 \label{eq:action}
\end{equation}
where $\dots$ stand for higher order corrections.  In the weak field limit $\C{R}\approx\Box\approx\C{F}^2\ll \mu^2\sim \mpl^2$, the functions $\mathfrak{F}$, $\mathfrak{L}$ match the conventional GR values, $\mathfrak{F}\to {\cal R}$ and $\mathfrak{L}\to {\cal F}^2$. The resulting field equations can be derived and, upon some regularity conditions for $\mathfrak{F}$ and $\mathfrak{L}$, can be cast in a form equivalent to Einstein's equations coupled to an effective energy momentum tensor
\be
\frac{1}{8\pi G}\left(\C{R}_{\mu\nu}-\frac{1}{2}g_{\mu\nu}\C{R}\right)= \C{T}_{\mu\nu}
\label{eq:einsteineqs}
\ee
where $\C{T}^{\mu\nu}=\C{T}_{\mathfrak{F}}^{\mu\nu}+\C{T}_\mathfrak{L}^{\mu\nu}$ contains both a gravity and an electromagnetic part. 

Accordingly the electromagnetic field equations can be written in a Maxwell's form, \textit{i.e.},
\be
\frac{1}{\sqrt{-g}}\, \p_\mu \left( \sqrt{-g} \C{F}^{\mu\nu}\right)=\C{J}_\mathfrak{L}^\mu
 \ee
where $\C{J}_\mathfrak{L}^\mu$ is an effective current depending on $\C{F}$, $\Box$ and $\mu$.
For specific profiles of the functions $\mathfrak{F}$ and $\mathfrak{L}$ one can consider examples in \cite{ANS07,MMN11,IMN13,Mod12a}.
 
Spacetime regularity implies conditions for  $m(r)$ and $q(r)$, namely
\bea
m(r)=O\left(r^3\right)\quad\mathrm{as}\quad r\to 0\non\\
q(r)=O\left(r^2\right)\quad\mathrm{as}\quad r\to 0.\non
\eea
Within the gravity ultraviolet self-complete paradigm, the above conditions are too restrictive. The Planck scale is dominated by a quantum pregeometry for which the metric description is just a loose concept. As a result, it is sufficient to require that the functions $m(r)$ and $q(r)$ exist in the interval $\left(\lpl, \infty\right)$. For the ease of discussion, we ask, however, $m(r)$ and $q(r)$ to be monotonically increasing functions and non-vanishing in the aforementioned interval. This is equivalent to excluding a  multi-horizon structure and having, at the most, two zeros for $F(r)$, namely an event horizon, $r_+$, and a Cauchy horizon, $r_-$, with $F(r_\pm)=0$.

From \eqref{eq:metric} one can calculate the black hole temperature that reads
\begin{equation}
T=\frac{1}{4\pi r_+}\left[1-r_+\frac{m^\prime(r_+)}{m(r_+)}-\frac{q^2(r_+)}{r_+^2}\left( 1+ r_+\frac{m^\prime(r_+)}{m(r_+)}-2r_+\frac{q^\prime(r_+)}{q(r_+)}\right)\right].
\label{eq:temperature}
\end{equation}
Being the function $m^\prime(r_+)=\left.\frac{\d m(r)}{\d r}\right|_{r=r_+}$ and $q^\prime(r_+)=\left.\frac{\d q(r)}{\d r}\right|_{r=r_+}$ positive defined, one obtains that there exists a zero temperature state corresponding to the extremal black hole configuration, $r_\mathrm{e} =r_-=r_+$. This is equivalent to saying that event horizon radii are limited from below, $r_+\geq r_\mathrm{e}$.

The metric \eqref{eq:metric} can describe a system that is simultaneously a black hole and an elementary particle if
\begin{equation}
r_+=\lambda
\label{eq:particlebh}
\end{equation}
where $\lambda$ is the Compton wavelength. The latter is expressed in terms of the black hole internal energy, namely
\begin{equation}
\lambda=\frac{2\pi}{M(r_+, Q)},\quad \mathrm{with}\quad M(r_+, Q)=\frac{1}{2 r_+\mathbb{H}(r_+) G}\left[r_+^2+q^2(r_+)\right],
\end{equation}
where $\mathbb{H}(r)=m(r)/M$. The temperature \eqref{eq:temperature} fulfills the first law of thermodynamics
\begin{equation}
\d M=T\d S+\Phi \d Q
\end{equation}
from which one obtains the area law $\d S=\d A_+ /4 G(r_+)$, being $\Phi$ the electrostatic potential. Here $G(r)=G\ \mathbb{H}(r)$ is an effective gravitational constant that takes into account for the higher derivative corrections to Einstein gravity in \eqref{eq:action}.

The particle-black hole equation \eqref{eq:particlebh} selects a family of parameters $(M_\lambda,\, r_\lambda)$ depending on $Q$. This is in contrast to what is found in the case of neutral black holes, whose particle-black hole equation is satisfied for a unique pair of black hole parameters. Such a presence of a family of solutions $(M_\lambda,\, r_\lambda)$ resembles the situation of regular neutral black hole solutions in the presence of a minimal length $\ell$ \cite{SpA11}. In the latter case, a unique pair of solutions has been identified by choosing the specific value of $\ell$ such that the particle-black hole condition is satisfied at the extremal configuration only, \textit{i.e.}, the smallest black hole. Such a procedure can be repeated in the case currently under investigation since for arbitrary values  $(M_\lambda,\, r_\lambda)$ the black hole might be rather unstable. Due to a nonvanishing temperature the hole  will decay and lose charge quickly. 
There are however two important differences with respect to the neutral regular black hole in \cite{SpA11}. First, the charge $Q$ cannot assume arbitrary values as $\ell$. By expressing the $Q$ in terms of the elementary charge, $Q=n\ e\ G^{1/2}$ with $n\in\mathbb{N}$, one obtains a natural quantization rule and a consequent pixelization of the event horizon. This is equivalent to saying that the equations
\begin{equation}
M_\lambda=M_\mathrm{e},\quad r_\lambda=r_\mathrm{e}\quad \Longrightarrow(M_0, r_0)
\end{equation}
might not have an exact solution but rather determine a quasi-extremal configuration $(M_0, r_0)$.
Second, if one assumes the decay of the deSitter space as a formation mechanism for charged Planckian black holes, it has to be noted in \cite{MaR95} that extremal black hole production rates are suppressed versus those of non-extremal ones. As a result, the quasi-extremal black hole $(M_0, r_0)$ would turn to be a favorite state despite having a non-vanishing temperature.


\section{Classicalization of charged black holes}
\label{sec:classicalization}

Up to now, our considerations have had a model independent character. We preferred not to specify the  profile of the functions $\mathfrak{F}$ and $\mathfrak{L}$ in \eqref{eq:action}. Such a choice is corroborated by the ultraviolet self-completeness scenario, according to which gravity is dominated by a classical state in the trans-Planckian regime. Only tiny modifications with respect to Einstein gravity are allowed around the Planck scale \cite{DFG11}. We will therefore neglect them in the present section by assuming the following conditions:
\begin{equation}
m^\prime\ll m/r, \quad q^\prime\ll q/r.
\label{eq:smallderivatives}
\end{equation}
As a result, the metric \eqref{eq:metric} actually describes a Reissner-Nordstr\"{o}m geometry up to subleading non-classical corrections.

The particle-black hole condition \eqref{eq:particlebh} therefore implies
\begin{equation}
M_\lambda =\frac{2\pi}{\sqrt{4\pi G-Q^2}}\, .
\label{eq:mlambda}
\end{equation}
From the above formula one obtains a bound for the charge $Q<\sqrt{4\pi G}$, which we assume positive defined for the ease of discussion. By plugging $M_\lambda$ in $r_+$ one obtains two values for $r_\lambda$. One of these coincides with the corresponding Compton wavelength 
and holds only for $Q<\sqrt{2\pi G}$. In such an interval, one has $r_\lambda>\lpl$.

By expressing $Q$ in terms of the elementary charge one can write \eqref{eq:mlambda} as
\begin{equation}
M_\lambda =\sqrt{\frac{\pi}{1-n^2\alpha/4\pi}}\ \mpl
\label{eq:mlambdafinale}
\end{equation}
where $\alpha=e^2$.
The bound on $Q$ implies a condition for the charge number $n$, namely $n\leq n_\mathrm{max}$, where
\begin{equation}
 n_\mathrm{max}=\left\lfloor \sqrt{\frac{2\pi}{\alpha}}\right\rfloor =29
 \label{eq:nmax}
\end{equation} 
Here $\lfloor x\rfloor$ indicates the floor function of $x$, $ {\displaystyle {\text{floor}}(x)=\lfloor x\rfloor } $, namely the greatest integer that is less or equal to $x$. In \eqref{eq:nmax} we have not taken into account the increase of the fine structure constant at the Planck scale. The Landau pole occurs at a scale
${\displaystyle \Lambda _{\text{Landau}}}\simeq 10^{286}$ eV $\gg \mpl$ but non-perturbative effects might change the scenario drastically. If the run to the Planck scale makes $\alpha$ of order unity, the value of $ n_\mathrm{max}$ would be reduced to $2$.   

Another bound on the charge is due to the electrostatic repulsion. If we consider a probe charge, \textit{e.g.} a proton, near a charged black hole, the gravitational attraction will overcome the electrostatic repulsion 
only if the hole charge $Q$ is bounded by the relation
\begin{equation}
Q<\left(\frac{G\mpr}{\qpr}\right) GM
\end{equation}
where $\mpr$ is the proton mass and $\qpr=G^{1/2}\ \sqrt{\alpha}$. The above relation, based on Newton's law and Coulomb's law arguments, is used to estimate the maximal charge a hole can accumulate \cite{Pag06}. By plugging $M_\lambda$ in the above inequality one gets
\begin{equation}
Q^2<2\pi G\left(1-\sqrt{1-\alpha^{-1}\left(\mpr/\mpl\right)^2}\right) GM\simeq \pi\alpha^{-1}\left(\mpr/\mpl\right)^2\, .
\end{equation}
 Being $\mpr/\mpl\simeq 7.68\times 10^{-20}$, particle black holes cannot accumulate charge unless they are nucleated in the early Universe with a given $Q$.

In order to fulfill the condition of extremal particle black hole one has to require $M_\lambda=M_\mathrm{e}$, where
\be
M_\mathrm{e}=n\sqrt{\alpha}\ \mpl.
\ee
As a result one finds $M_\mathrm{e}=\sqrt{2\pi}\mpl$ and $r_\mathrm{e}=Q_\mathrm{e}=\sqrt{2\pi}\lpl$. Clearly this condition cannot be met since the electric charge is quantized. The closest configuration to the extremal one has a charge $Q_0=\left\lfloor Q_\mathrm{e}\right\rfloor$. Not surprisingly $Q_0$ corresponds to a charge number, $n_0=n_\mathrm{max}$.  The corresponding horizon radius and black hole mass are
\bea 
r_0&=&\sqrt{4\pi-n_0^2\alpha}\ \lpl\simeq 2.54\ \lpl\\
M_0&=&\frac{2\pi}{\sqrt{4\pi-n_0^2\alpha}}\ \mpl\simeq 2.48\ \mpl \, .
\eea
From this perspective $r_0$ represents the minimal horizon radius for a black hole mass $M_0$. For comparison, the neutral particle-black hole, $n=0$, has a radius $r_\mathrm{S}\simeq 3.54 \lpl$ and a mass $M_\mathrm{S}\simeq 1.77 \mpl$.

The quasi-extremality condition implies a non-vanishing temperature, namely
\be
T_0=\frac{\mpl}{4\pi\sqrt{4\pi-n_0^2\alpha}}\left[1-\frac{n_0^2\alpha}{4\pi-n_0^2\alpha}\right] \simeq \frac{\mpl}{2\pi\sqrt{2\pi}}\left(1-\frac{n_0\alpha}{2\pi}\right)
\label{eq:quasiexttemp}
\ee
that corresponds to $T_0\simeq 1.48\times 10^{-3}\ \mpl$ or equivalently $T_0\simeq 2.09\times 10^{29}$ K. Such black holes would be very hot and bright but they are not expected to suffer from relevant quantum back reaction being $T_0/M_0\simeq 5.98\times 10^{-4}$. In contrast to Planckian neutral black holes, they can safely be described in terms of semiclassical gravity, despite their minuscule size. Another difference with respect to the neutral case is the sign of the heat capacity
\be
C = 2\pi\ \frac{r_+^2-Q^2}{\left(3Q^2r_+^{-2}-1\right)}.
\ee
Being $Q_\mathrm{e}\lesssim r_0< \sqrt{3}Q_\mathrm{e}$, the heat capacity is positive.
As a result charged quasi-extremal particle-black holes enjoy thermodynamic stability.

For the entropy one obtains
\be
S_0=\pi \left(4\pi-n_0^2 \alpha\right) \simeq 20.2\, .
\ee
Being the production rate of extremal black holes suppressed relative to quasi-estremal ones by a factor $e^{S_0}\simeq 5.92 \times 10^{8}$, the production of such charged particle-black holes in the early Universe would be highly favored with respect to short scale modified neutral black hole remnants with a degenerate horizon \cite{NSS06b,MaN11,SpA11,NiS12}.

\section{Planckian charged black hole decay}
\label{sec:decay}

The main result of the previous section is that the presence of charge prevents a clear separation between particles and black holes by means of a stable particle-black hole state at the Planck scale. The presence of the charge allows for quasi-extremal particle-black holes at the best. A natural question to address concerns now the nature of their evolution. 

The Hawking temperature \eqref{eq:quasiexttemp} implies a rapid decay that is accompanied by a loss of mass and charge. If we now consider the emission of a positron $(\mel, |e|)$, the charge  variation is $\delta Q_0=-|\delta Q_0|= -\sqrt{\alpha}\ \lpl$, while the mass variation
\begin{equation}
\delta M_0=-|\delta M_0|= -\mel +\frac{2\pi Q_0\ \delta Q_0}{\left(4\pi G-Q_0^2\right)^{3/2}}\ \simeq -\frac{2\pi n_0\ \alpha}{\left(4\pi-n_0^2\ \alpha\right)^{3/2}}\ \mpl
\end{equation}
is controlled by the charge variation being $\mel\ll \mpl$.
From the comparison of the first order corrections to the Compton wavelength and the horizon radius, 
\begin{eqnarray}
&&\frac{2\pi}{M_0}\left(1+\frac{|\delta M_0|}{M_0}\right)\non\\ 
&&\neq GM_0\left(1-\frac{|\delta M_0|}{M_0}\right)\left[1+\sqrt{1-\frac{Q_0^2}{G^2 M_0^2}\left(1-2\ \frac{|\delta Q_0|}{Q_0}+2\ \frac{|\delta M_0|}{M_0}\right)}\right],
\end{eqnarray}
it turns out that the evaporation will disrupt the particle-black hole condition. Being $|\delta Q_0|/Q_0\simeq |\delta M_0|/M_0$ and $Q_0^2/(G^2 M_0^2)\simeq 1$,   the Compton wavelength increases while the horizon radius decreases. This means that the state $(r_0, M_0)$ will not evolve to a state $(r_\lambda, M_\lambda)$ but rather to a generic quasi-extremal  charged black hole configuration $(r_+, M)$. From here on, such a black hole will keep radiating towards a neutral or quasi neutral black hole configuration. In the process the black hole will progressively decrease its temperature. In such a phase the condition \eqref{eq:smallderivatives} cannot longer be valid and the temperature will be described by \eqref{eq:temperature}. The destiny of the black hole could be either an extremal neutral  particle-black hole configuration or an extremal charged black hole configuration. We underline here that the presence of charge would prevent an extremal particle-black hole configuration for what discussed in the previous section. This would represent a serious limitation to the gravity self-completeness paradigm when the charge is taken into account.

To address this ambiguity on the final fate of quasi-extremal charged particle-black hole we recall that the Hawking emission is accompanied by another emission process, namely the Schwinger $e^\pm$ pair production.  The latter occurs for the presence of intense electric fields, $E\geq E_\mathrm{c}=\pi \mel^2/|e|$,  near the event horizon. In our case the condition leads to
 \begin{equation}
 \frac{Q}{r^2_+}\ \mpl\ge \pi\, \frac{m^2_e}{\sqrt{\alpha}} \label{eq:schwinger}
\end{equation}
Being $r_+\sim \lpl$,  one has that the field at the horizon is overcritical, $E(r_+)\geq E_\mathrm{c}$, if $n\geq (\pi/\alpha)(\mel/\mpl)^2$.  This means that it is sufficient to have just one elementary charge to trigger the $e^\pm$  pair production at the event horizon. One particle of the pair will be rejected while the other absorbed leading to a quick hole's discharge. Such a scenario is in agreement with early findings about the spontaneous loss of charge of black holes \cite{Gib75}. 

In conclusion the final stages of the evolution of the quasi-extremal particle-black hole will be described by a neutral configuration asymptotically relaxing to a zero temperature stable particle-black hole state.


As a final comment we want to address particle black holes in the opposite regime, \textit{i.e.},  quasi-neutral particle black holes $(r_\lambda, M_\lambda)$, one can obtain for $n\sim 1$. From \eqref{eq:mlambdafinale} one obtains $M_\lambda\simeq M_\mathrm{S}$ and $r_\lambda\simeq r_\mathrm{S}$, being $\alpha/4\pi\ll 1$. From the calculation of their entropy one can estimate the production rate in the early Universe relative to extremal black holes. It turns out that the relative rate is $1.37\times 10^{17}$, making them even more probable deSitter space decay products with respect to the quasi-extremal particle-black holes.  This is in agreement to the fact that they have lighter masses. As mentioned in Section \ref{sec:chargedmetric}, they decay more wildly since their temperature $T\sim 2.24\times 10^{-2}\ \mpl\simeq 3.19\times 10^{30}$ K is hotter  than \eqref{eq:quasiexttemp}. The Hawking emission will be accompanied by a Schwinger discharge. As a result, we expect a quick formation of a neutral black hole configuration. The latter might approach a stable zero temperature particle-black hole configuration provided the subleading corrections of the function $m(r)$ are taken into account.
  
\section{Conclusions} 

\label{sec:concl}


In this paper we analyze the role of the charge within the self-complete gravity paradigm. We presented the thermodynamics for a general short scale modified gravity Planckian black hole solution. We showed that the particle-black hole condition is satisfied by a family of black hole parameters. By assuming an approximately classical black hole state, we showed that the charge prevents the formation of zero temperature particle-black hole configuration. Only quasi-extremal particle-black holes are allowed, that could have  plentifully been produced during the early Universe quantum mechanical decay. We showed that, due to their temperature, they can decay in lighter black holes but not in particle-black holes. We also studied the Schwinger pair production mechanism and we concluded that at the event horizon the electric field is always overcritical provided one elementary charge is left on the hole. This fact allowed to solve the potential ambiguity about the fate of such quasi-extremal particle-black holes. Rather than cooling down to a extremal charged configuration, they will slowly relax towards a neutral configuration. We also showed that the latter can fulfill the condition of extremal particle-black hole provided corrections to the classical metrics are taken into account.

We also commented about the evolution of quasi-neutral particle-black holes, by showing that they will rapidly decay via Hawking and Schwinger mechanisms towards an extremal neutral particle-black hole configuration.

In conclusion, we have shown that the charge seriously disturbs the stability of the particle-black hole phase diagram at the Planck scale. The self-completeness is however safely restored thanks to a rapid discharge the charged particle-black hole is expected to undergo after its formation.

\subsection*{Acknowledgments}

The work of P.N. has been supported by the project ``Evaporation of the microscopic black holes'' of the German Research 
Foundation (DFG) under the grant NI 1282/2-2.

\end{document}